\documentclass[twocolumn,pra,showpacs,superscriptaddress]{revtex4-1}
\usepackage{graphicx}
\usepackage{subfigure}
\usepackage{amsmath}
\usepackage{amsfonts,amssymb}
\usepackage{bm}
\usepackage{float}
\usepackage{color}
\usepackage{sidecap}
\allowdisplaybreaks
\usepackage{soul}
\setstcolor{red}
\usepackage{ulem}
\usepackage{hyperref}
\usepackage[title,titletoc,header]{appendix}
\hypersetup{pdfpagemode=FullScreen,colorlinks=true,breaklinks,urlcolor=blue,linkcolor=blue,citecolor=blue}

\begin{document}

\title{Quench dynamics in the Aubry-Andr\'e-Harper model with \textit{p}-wave superconductivity}

\author{Qi-Bo Zeng}
\email{zqb15@mails.tsinghua.edu.cn}
\affiliation{Department of Physics and State Key Laboratory of Low-Dimensional Quantum Physics, Tsinghua University, Beijing 100084, China}

\author{Shu Chen}
\affiliation{Beijing National Laboratory for Condensed Matter Physics, Institute of Physics,
Chinese Academy of Sciences, Beijing 100190, China}
\affiliation{Collaborative Innovation Center of Quantum Matter, Beijing, China}

\author{Rong L\"u}
\email{rlu@tsinghua.edu.cn}
\affiliation{Department of Physics and State Key Laboratory of Low-Dimensional Quantum Physics, Tsinghua University, Beijing 100084, China}
\affiliation{Collaborative Innovation Center of Quantum Matter, Beijing, China}

\begin{abstract}
The Anderson localization phase transition in the Aubry-Andr\'e-Harper (AAH) model with \textit{p}-wave superconducting (SC) pairing is numerically investigated by suddenly changing the on-site potential from zero to various finite values which fall into the extended, critical and localized phase regimes shown in this model. The time evolutions of entanglement entropy (EE), mean width of wave packets and Loschmidt echo of the system exhibit distinct but consistent dynamical signatures in those three phases. Specifically, the EE grows as a power function of time with the exponent of which varies in the extended phase but keeps almost unchanged in the critical phase for different quench parameters. However, if the system is in the localized phase after a quench, the EE grows much slower and will soon get saturated. The time-dependent width of wave packets in the system shows similar behaviors as the EE. In addition, from the perspective of dynamical phase transition, we find that the Loschmidt echo oscillates and always keeps finite when the system is quenched in the extended phase. In contrast, in the critical or localized phase, the echo will reach to zero at some time intervals or will decay almost to zero after a long-time evolution. The universal features of these quantities in the critical phase of the system with various SC pairing amplitudes are also observed.
\end{abstract}


\maketitle
\date{today}

\section{Introduction}

The Aubry-Andr\'e-Harper (AAH) model has attracted considerable attentions both theoretically and experimentally in the past few decades \cite{Harper, Aubry, Hiramoto1, Ostlund, Kohmoto1, Kohmoto2, Thouless, Hiramoto2, Geisel, Han, Chang, Takada, Liu, Wang}. With the fast development of experimental technologies and skills, this model can now be realized in photonic crystals \cite{Negro, Lahini, Kraus1} and cold-atom systems \cite{Roati, Modugno}, which help us to gain more understandings of incommensurate systems. With so many important and interesting properties shown in this one dimensional incommensurate model, the Anderson localization phase transition has been one of the most extensively explored phenomena. Besides, recent studies on the many-body localization phase transition in the AAH model with interactions between the neighboring sites endows such systems with more exciting features \cite{Iyer, Schreiber, Li, Iemini}. Moreover, many generalized AAH models with exotic characteristics have been proposed, such as the AAH model with \textit{p}-wave superconducting (SC) pairing terms \cite{Wang, Zeng, HQWang}. Due to the SC pairing, this system shows a critical region before the transition from extended phase into the localized phase \cite{Wang}. However, the phase transition process, especially the critical region has not been deeply discussed. Many features of the critical phase still need to be clarified, which is one of the purposes of this paper.

On the other hand, the quantum entanglement is becoming more and more important in the research of many-body theory and quantum information theory \cite{Amico, Horodecki, Eisert}. By checking the entanglement of the system, it has been found that the behaviors of entanglement are closely related to phase transitions \cite{Osborne, Osterloh, LAWu, Alcaraz, Xavier}. The features in the entanglement for critical systems are quite distinctive and thus can be exploited to study the phase transitions. Entanglement entropy (EE) \cite{Eisert}, as a measure of the entanglement, plays a central role in characterizing different phases. By studying the variations and scalings of the EE at or near the critical points, one can get more insight in the phase transition process. In Ref. \cite{Shem, Igloi, Roosz}, the EE and the entanglement spectra of an Aubry-Andr\'e model are elucidated and clear signatures of the Anderson localization phase transition have been obtained. If the \textit{p}-wave superconducting pairing term is added to this model, one could expect more interesting phenomena due to the interplay between the SC pairing and the Anderson localization. Furthermore, the behaviors of EE in the critical region of the AAH model with SC pairing could also be studied, from which more interesting features of the phase transitions in this model can be revealed. It is also noteworthy that nowadays the quantum quench method has become very effective in the study of dynamical quantum phase transitions (DQPTs) \cite{Heyl1, Karrasch, Canovi, Heyl2, Heyl3, Huang}. The Loschmidt echo, which can be used to characterize the overlap between the initial state and the time-evolved state after a quench, is widely utilized to describe the DQPTs \cite{Jalabert, Gorin, Quan, Jafari}. It is known that the appearance of zeros in the Loschmidt echo signifies the dynamical phase transitions. In Ref. \cite{Yang}, the Loschmidt echo is used to describe the localization-delocalization phase transition in AAH model and distinct signatures of the phase transition are acquired. It would be interesting to further investigate the influence of the \textit{p}-wave SC pairing on the Loschmidt echo behaviors in similar models.

\begin{figure}[!ht]
\centering
\includegraphics[width=3.6in]{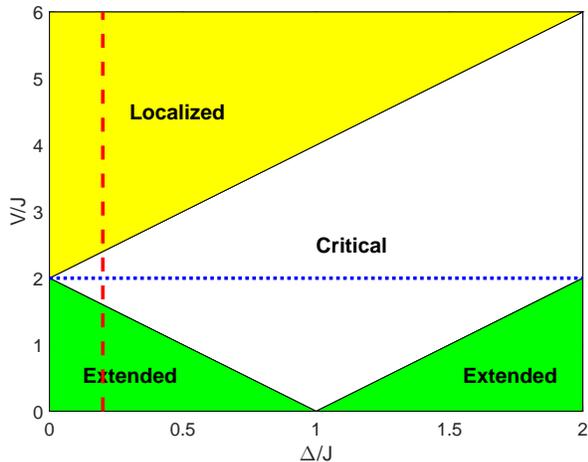}
\caption{(Color online)  Phase diagram of the AAH model with p-wave superconducting pairing. Three different phases show up in different parameter regimes: extended, critical and localized phase. Here $\Delta$ is the p-wave SC pairing amplitude, $J$ is the nearest hopping amplitude and $V$ is the on-site potential, see Eq. (\ref{Eq1}). The red dashed line corresponds to $\Delta=0.2J$, where we do the quench along this line by varying $V$ from zero to finite values fall into different phase regimes. The blue dotted line shows the quench in the critical region by changing the SC pairing amplitude at $V=2J$.}
\label{fig1}
\end{figure}

Motivated by these results, in this paper, we study a generalized AAH model with \textit{p}-wave superfluidity from the perspective of quench dynamics. As we already know that such a system would show extended, critical and localized phases in different parameter regimes. The phase diagram of the system is presented in Fig. \ref{fig1}, which is similar to the one in Ref. \cite{Wang}. To investigate the dynamical properties of these three phases, the system is initially prepared with the on-site potential set to be zero. Then we suddenly change the on-site potential to various finite values which fall into the extended, critical and localized region respectively. After that, the time evolutions of entanglement entropy, mean width of wave packets and Loschmidt echo in this system are checked. Distinct signatures are observed for the system quenched into different phases. As to the entanglement entropy, if the system is in the extended and critical phase after a quench, it grows as a power function of time. The exponent of the power function varies in the extended phase but keeps almost unchanged in the critical phase for different quench parameters. However, if the system is quenched into the localized state, the EE increases quite slowly and will soon become saturated. Similar behaviors also show up in the evolution of the mean width of wave packets. Furthermore, we also explored the Loschmidt echo of the system from the perspective of dynamical phase transitions. The Loschmidt echo oscillates at some relatively large values and always keeps finite when the system is in the extended state after the quench, but will approach to zero at some time intervals when the system goes into the critical regime. If we quench the system into the localized phase, the Loschmidt echo decays almost to zero in the long-time evolution. Besides, the variations of these quantities in the critical phase of system with various SC pairing amplitudes are also discussed, where the universal properties are also analyzed. The dynamical signatures obtained in this work are important in characterizing the non-equilibrium properties of the different phases in the AAH model with \textit{p}-wave SC pairing.

The rest of the paper is organized as follows. In Sec. \ref{sec2}, we introduce the model Hamiltonian of the AAH model with \textit{p}-wave superconducting pairing. Then in Sec. \ref{sec3}, we discuss the time evolution of entanglement entropy, mean width of wave packets  and Loschmidt echo in the system after a quantum quench. Different quench situations are investigated and the corresponding numerical results are presented. Sec. \ref{sec4} is dedicated to a brief summary.

\section{Model Hamiltonian}\label{sec2}

The generalized one-dimensional (1D) Aubry-Andr\'e-Harper model with \textit{p}-wave superconducting (SC) pairing we consider in this paper is described by the following Hamiltonian

\begin{equation}\label{Eq1}
H = \sum_{j=1}^{N} V_j c_j^\dagger c_j + \sum_{j=1}^{N-1} [ -J c_{j+1}^\dagger c_j +\Delta c_{j+1}^\dagger c_j^\dagger + H.c. ],
\end{equation}
where $c_j^\dagger$ ($c_j$) is the creation (annihilation) operator at site $j$, $V_{j} = V \cos(2\pi \alpha j)$ is the on-site potential and $\alpha$ is set to be $(\sqrt{5}-1)/2$ without loss of generality. $J$ is the hopping amplitude between the nearest neighbouring lattice sites and we set $J=1$ as the energy unit throughout this paper. $\Delta$ is the superconducting pairing amplitude which is taken to be real. $H.c.$ represents the Hermitian conjugate. There are in total $N$ lattice sites in this system. It has been shown that when $V$ is larger than a certain critical value, $V > 2(J+\Delta)$, the system will go through a phase transition from the extended state into the localized state. More recently, it is found that with nonzero SC pairing, there will be a critical phase before the system becomes localized \cite{Wang, Zeng}. It will be interesting and helpful to investigate this phase transition from the perspective of non-equilibrium dynamics. The method we employed here is to check the behaviors of the physical quantities of the system after a global quench, namely a sudden change of the system parameters. In order to calculate the quantities we will discuss later, first we need to rewrite the Hamiltonian by using the Bogoliubov-de Gennes (BdG) transformation:
\begin{equation}
\eta_n^\dagger = \sum_{j=1}^{N} [ u_{n,j} c_j^\dagger + v_{n,j} c_j ],
\end{equation}
where $n=1,...,N$. The $u_{n,j}$ and $v_{n,j}$ are chosen to be real here. Then the Hamiltonian in Eq. (\ref{Eq1}) can be diagonalized as
\begin{equation}
H = \sum_{n=1}^{N} \epsilon_n (\eta_n^\dagger \eta_n - \frac{1}{2})
\end{equation}
with $\epsilon_n$ being the energy of quasiparticles. For the components of $u_{n,j}$ and $v_{n,j}$, we have the following BdG equations:
\begin{widetext}
\begin{equation}
\left\{
  \begin{aligned}
  -J u_{n,j-1} + \Delta v_{n,j-1} + V_j u_{n,j} - J u_{n,j+1} - \Delta v_{n,j+1} &= \epsilon_n u_{n,j},\\
  -\Delta u_{n,j-1} + J v_{n,j-1} - V_j v_{n,j} + \Delta u_{n,j+1} + J v_{n,j+1} &= \epsilon_n v_{n,j}.
  \end{aligned}
 \right.
\end{equation}
\end{widetext}
Representing the wave function as
\begin{equation}\label{}
  | \Psi_n \rangle = [u_{n,1}, v_{n,1}, u_{n,2}, v_{n,2}, \cdots, u_{n,N}, v_{n,N}]^T,
\end{equation}
the BdG equations can be written as $\mathcal{H}  | \Psi_n \rangle = \epsilon_n  | \Psi_n \rangle$, with $\mathcal{H}$ being a $2N \times 2N$ matrix:
\begin{equation}\label{Eq6}
\mathcal{H}=
  \begin{pmatrix}
    A_1 & B & 0 & \cdots & \cdots & \cdots & 0 \\
    B^\dagger & A_2 & B & 0 & \cdots & \cdots & 0 \\
    0 & B^\dagger & A_3 & B & 0 & \cdots & 0 \\
    \vdots & \ddots & \ddots & \ddots & \ddots & \ddots & \vdots \\
    0 & \cdots & 0 & B^\dagger & A_{N-2} & B & 0 \\
    0 & \cdots & \cdots & 0 & B^\dagger & A_{N-1} & B \\
    0 & \cdots & \cdots & \cdots & 0 & B^\dagger & A_N
  \end{pmatrix},
\end{equation}
where
\begin{equation}\label{}
  A_j =
  \begin{pmatrix}
    V_j & 0 \\
    0 & -V_j
  \end{pmatrix},
\end{equation}
\begin{equation}\label{}
  B=\begin{pmatrix}
      -J & -\Delta \\
      \Delta & J
    \end{pmatrix},
\end{equation}
The energy spectrum as well as the eigenvectors can be determined by diagonalizing this matrix directly. The ground state of the system corresponds to the state with all negative quasiparticle energy levels filled.

In the next section, we will numerically explore the quench dynamics of entanglement entropy, mean width of wave packets and Loschmidt echo of this incommensurate AAH model. The dynamical behaviors of these quantities after quench will be shown and analyzed.

\section{Numerical results and discussions}\label{sec3}

In this section, we will study the non-equilibrium properties of the Anderson localization phase transition of the AAH model by exploiting the quantum quench method. Firstly we will discuss the variation of the entanglement entropy after we quench the system into different phases. Then the time evolutions of mean width of wave packets and Loschmidt echo in this system will also be studied. The comparison between these physical quantities will also be presented.

\subsection{Entanglement entropy}

The entanglement entropy (EE), which is also called the von-Neumann entropy, is defined to quantify the entanglement of a system. The EE can be calculated firstly by dividing the whole Hilbert space into two subspaces A and B. Suppose that $| \Omega \rangle$ is the quantum state and $\rho = | \Omega \rangle \langle \Omega |$ is the density matrix for the full system. The reduced density matrix for subsystem A can be obtained by tracing out the degrees of freedom of subspace B, which leads to
\begin{equation}
\rho_A = Tr_B | \Omega \rangle \langle \Omega |.
\end{equation}
Then the entanglement entropy is defined as
\begin{equation}
S_A = -Tr_A ( \rho_A \log \rho_A )
\end{equation}

According to Refs. \cite{Peshel1, Peshel2}, the entanglement entropy for non-interacting systems can be calculated from the correlation functions. For the quadratic fermionic system we discuss here, the time evolution of the entanglement entropy can be calculated by using the method which is detailed described in the Appendix A of Ref. \cite{lgloi2}. Before the quench, we set the system parameters such as the on-site potential or the superconducting pairing amplitude to be certain initial values. Then at $t=0$, we suddenly change the parameters to values different from the initial ones.

\begin{figure}[!ht]
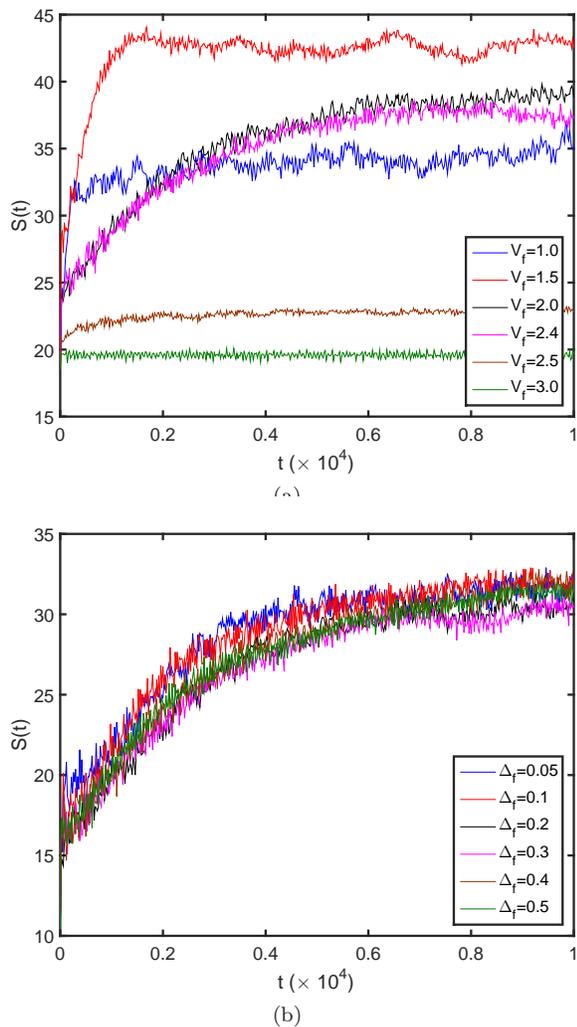

\centering
\subfigure[]{
\label{fig2a}
\includegraphics[width=3.0in]{fig2a.eps}}
\subfigure[]{
\label{fig2b}
\includegraphics[width=3.0in]{fig2b.eps}}
\caption{(Color online) The time evolution of the entanglement entropy in the AAH model with \textit{p}-wave superfluidity after a quench. (a) The on-site potential in the model Hamiltonian is set to be 0 before the quench and then it is suddenly changed to different $V_f$s with $1.0 \leq V_f \leq 3.0$ at time $t=0$. Here the SC pairing amplitude is fixed at $\Delta=0.2$. (b) The initial SC pairing amplitude $\Delta_i=0.001$ and then is changed to a different $\Delta_f$ within the range $0.05 \leq \Delta_f \leq 0.5$. The on-site potential is $V=2.0$, so the system is always in the critical regime. The number of the lattice sites is chosen to be $N=233$, which is the 13th Fibonacci number.}
\label{fig2}
\end{figure}

Now let us investigate the time-evolution of the system's entanglement entropy. It is known that due to the existence of SC pairing, the system is in the extended phase when $V < 2|J-\Delta|$ and in the localized phase when $V > 2|J+\Delta|$. While if $2|J-\Delta| < V < 2|J+\Delta|$, the system will be in the critical phase (see Fig. \ref{fig1}). Since the system shows three different phases in different parameter regimes of disordered potential $V$, we will do the quench by setting $V_i=0$ in the initial Hamiltonian, which means that the system is prepared in the extended state. Then at time $t=0$, we suddenly change $V$ to different nonzero values, i.e. $V_f$s, as shown by the red dashed line in Fig. \ref{fig1}. The variation of EE can be calculated as the system evolves after the quench.

In Figure \ref{fig2}, the time-dependent EE is presented for different quench parameters. We choose $J=1$ as the energy unit of the system throughout this paper. The lattice number of the system is $N=233$, which is the 13th Fibonacci number. After the quench, we can see that the EE grows as a power function of time, namely, we have
\begin{equation}\label{}
S(t) \sim t^\sigma.
\end{equation}
However, the specific dynamical behavior for different $V_f$s changes. To better characterize the power-law form of the growth profile, we can plot the EE as a function of time in a log-log coordinate systems, see Fig. \ref{figA1} in the Appendix. The linear growth part in that figure indicates that the EE do grows in a power-law form. If the system is quenched into the extended state ($V_f=1.0$, $1.5$ in Fig. \ref{fig2a}), the EE will increase quickly in the beginning. For example, when $V_f=1.5$, the exponent $\sigma \approx 0.23$, as shown by the red dashed line in Fig. \ref{figA1}(a). The exponent $\sigma$ becomes smaller as we further increase $V_f$. The EE will finally saturate to a finite value $\bar{S} \sim N$. If the quench is performed to the critical phase ($V_f=2.0$, $2.4$), the EE will grow continuously in a power-law form for a relatively longer time. The exponents for those power-law growth of EE are almost the same as long as $V_f$ falls into the critical region. From the numerical fitting in Fig. \ref{figA1}(a), we have $\sigma \approx 0.18$ for the critical phase (see the black dashed line there). The unchanged exponent characterizes the dynamical behaviors of EE in the critical phase. When the system is quenched to the localized phase ($V_f=2.5$, $3.0$ in Fig. \ref{fig2a}), the EE after a quench grows much slower and the corresponding exponent of the power function will be significantly reduced if the on-site potential becomes stronger. In the long-time evolution, we can see that the EE will always get saturated. The saturation of EE differentiates the Anderson localized phase from the many-body localized phase,  which has a unbounded logarithm growth with time \cite{Znidaric, Bardarson}. These distinctive behaviors of the EE in the quench dynamics clearly exhibit the different phases of the AAH system with p-wave SC pairing.

On the other hand, we could also check the evolution of the EE after suddenly changing the SC pairing amplitude of the system, as shown by the blue dotted line in Fig. \ref{fig1}. Such a quantum quench is helpful in gaining more insights about how SC pairing would affect the system's properties. Here we fix the on-site potential $2.0$, thus the system would always be in the critical phase. The numerical results are presented in Fig. \ref{fig2b}. The SC pairing amplitude of the initial Hamiltonian is chosen to be a small value, $\Delta_i=0.001$. Then it is suddenly changed to values ranging from $0.05$ to $0.5$. From Fig. \ref{fig2b} we can see that, even though the SC pairing amplitudes after the quench are quite different, the EE grows in a power-law form and the exponents for these power functions are almost the same. The numerical fitting in Fig. \ref{figA1}(b) shows that the value $\sigma$ is about 0.21 in these cases. The similar behaviors again demonstrate the signature the critical phase.

\subsection{Wave packet dynamics}

To further explore the dynamics of system, we now turn to study the time-evolution of wave packets. Here we mainly focus on the evolution of the mean width of wave packets. Following the previous studies \cite{lgloi1,lgloi2}, we suppose that at $t=0$, the following wave packets connecting site $k$ and $k^\prime$ can be constructed
\begin{widetext}
\begin{equation}\label{}
  W_{k,k^\prime}(t) = \frac{1}{2}\sum_n \{ \cos(\epsilon_n t) [ u_{n,k} u_{n,k^\prime} + v_{n,k} v_{n,k^\prime} ] - i \sin(\epsilon_n t) [ u_{n,k} v_{n,k^\prime} + u_{n,k^\prime} v_{n,k} ] \},
\end{equation}
\end{widetext}
where $\epsilon_n$ and $v_{n,k}$ are the eigenvalues and eigenvectors of the system Hamiltonian, which can be determined by diagonalizing the corresponding Hamiltonian matrix shown in Eq. \ref{Eq6}. $k$ and $k^\prime$ correspond to different sites. The width of the wave packet at site $k$ at time $t$ is
\begin{equation}\label{}
  d(k,t) =  \sqrt{ \sum_{k^\prime} (k-k^\prime)^2 |W_{k,k^\prime}|^2 }.
\end{equation}
We then take the average of $d(k,t)$ over the starting positions k, and get the mean width of wave packets in the system at time $t$ as
\begin{equation}\label{}
  d(t) = \frac{1}{N} \sum_k d(k,t).
\end{equation}
Since the system can be in the extended, critical or localized phase, we need to check whether the propagation of wave packets will be different in these three different phases.

\begin{figure}[!ht]
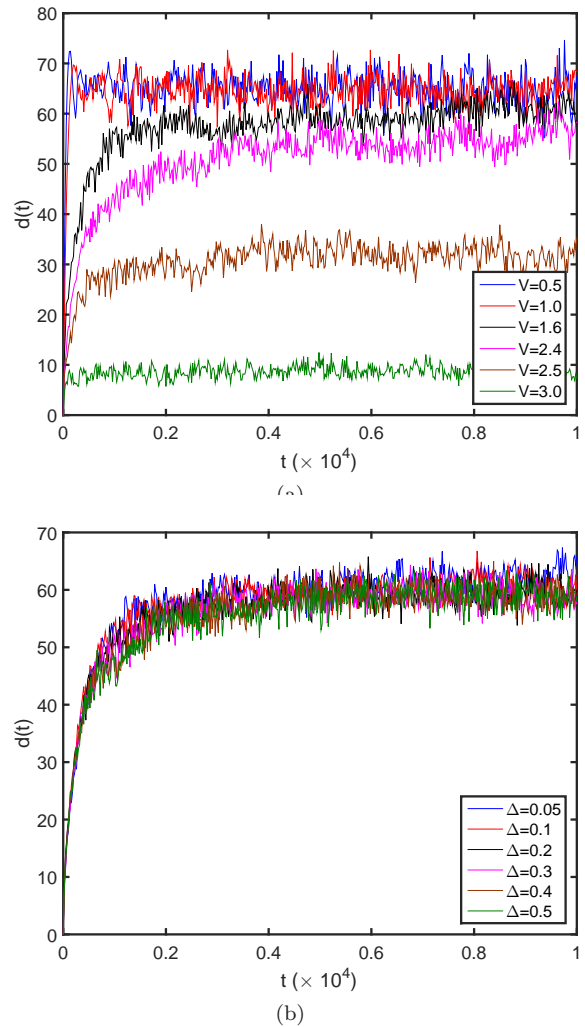

\centering
\subfigure[]{
\label{fig3a}
\includegraphics[width=3.0in]{fig3a.eps}}
\subfigure[]{
\label{fig3b}
\includegraphics[width=3.0in]{fig3b.eps}}
\caption{(Color online) Time-dependent width of wave packets in the AAH model with \textit{p}-wave SC pairing. (a) Mean width of wave packets in the systems with different on-site potentials $V=0.5- 3.0$. The SC pairing amplitude is $\Delta=0.2$. (b) Mean width of wave packets in the systems with various SC pairing amplitudes, $\Delta=0.05-0.5$. Here the on-site potential is fixed to $V=2.0$, so the system is always in the critical phase.  The number of the lattice sites is $N=233$.}
\label{fig3}
\end{figure}

In Figure \ref{fig3}, the time-dependent widths of wave packets in the AAH model with different on-site potentials and SC pairings are presented. From Fig. \ref{fig3a}, we can find that the average width of wave packets also grows in a power-law form after the quench:
\begin{equation}\label{}
  d(t) \sim t^D.
\end{equation}
The growth profile changes for different quench parameters. We also show the time evolution of $d(t)$ in the log-log frame in Fig. \ref{figA2}. The numerical results clearly show that when the system is in the extended state after a quench, see the blue and red line with $V=0.5$ and $1.0$ in Fig. \ref{fig3a}, the mean width of wave packets will increase quickly first and then saturate. Numerical fitting in Fig. \ref{figA2}(a) indicates that $D \approx 0.71$ for the $V=0.5$ cases (see the blue dashed line there). The growth behaviors of the width of wave packet is quite similar to that in the entanglement entropy (see Fig. \ref{fig2a}). This can be explained from the perspective of quasiparticles. The wave packets can be used to semiclassically describe the quasiparticles which are assumed to be produced uniformly in the system and will move classically after production. Since the system we discussed here is quasiperiodic, the quasiparticle excitations at time $t=0$ can only propagate diffusively ($x \sim t^D$ with $0<D<1$, where $x$ is the distance the quasiparticle travels during the time interval $[0,t]$), not like the ones in the homogeneous lattice which could move ballistically ($x \sim t$) \cite{lgloi1}. So the width of wave packets grows in a power-law form. The expansion of wave packets will lead to a growth of the entanglement entropy because these wave packets will become overlapped after expanding for a while and make the system more entangled. The same reason applies to the critical phase where the width of wave packets and EE also grows as a power function of time, as shown by the lines corresponding to $V = 1.6$ and $2.4$ in Fig. \ref{fig3a}. The exponents for the power functions in the critical phase are the same: $D \approx 0.47$ (see the black dashed line in Fig. \ref{figA2}(a)). So the universal feature of the critical phase is also exhibited in the dynamical behavior of wave packets. On the contrary, if the system is in the localized phase (see the lines with $V=2.5$ and $3.0$ in Fig. \ref{fig3a}), the wave packets will be confined near the original sites, so the time-dependent width increases much slower and will soon saturate to a finite value in the end. The exponent for the power function becomes smaller as $V$ is further increased. Now that the wave packets do not expand that much, they will be much less overlapped, which results in a system with small entanglement entropy. This is again consistent with the dynamical features of EE in the system after a quench to the localized phase.

The variation of $d(t)$ for systems with various SC pairing amplitudes is also analyzed here. The numerical results are presented in Fig. \ref{fig3b}. The on-site potential is also chosen to be $2.0$, so the system would always be in the critical phase. When $\Delta$ becomes larger, the critical region of the system will expand. However, the growth behaviors of the corresponding mean width of wave packets are almost the same. They all grow as a power function of time before getting saturated and the exponents of these power functions are very close. From Fig. \ref{figA2}(b), we have $D \approx 0.45$. This is again analogous to the behaviors of EE, as shown in Fig. \ref{fig2b}.

From the above discussions on the dynamical features of EE and mean width of wave packets, we can find that these two quantities behave quite similarly in the post-quench evolution even though the exponents do not match. The reason behind this is that the expansion of wave packets or the motion of quasi-particles would result in variations of the entanglement in the system, as discussed in Ref. \cite{lgloi1}.

\subsection{Loschmidt echo}

We can also use the Loschmidt echo to characterize the quench dynamics in the system. Loschmidt echo has been widely exploited in studying the dynamical quantum phase transitions. It is well known that when the system is going through a dynamical phase transition, the Loschmidt will approach to zero at some time intervals. In Ref. \cite{Yang}, the Loschmidt echo has been utilized to characterize the localization-delocalization phase transition in the AAH model, which shows clear signatures in different phases. For the AAH model with \textit{p}-wave SC pairing we discuss here, we can also check how the Loschmidt echo would behave. Due to the presence of SC pairing, a critical region exists, from which more interesting features could be expected. To define the Loschmidt echo, one can assume that the system is prepared in an eigenstate of the initial Hamiltonian $H(\lambda_i)$, where $\lambda_i$ could be the on-site potential $V_i$ or the SC pairing amplitude $\Delta_i$ before the quench. At $t=0$ we change $\lambda_i$ suddenly to another value $\lambda_f$, and let the system evolve under the new Hamiltonian $H(\lambda_f)$. The Loschmidt amplitude (also called the return amplitude) is defined as
\begin{equation}\label{}
  G(t, \lambda_i, \lambda_f) = \langle \psi(\lambda_i) | e^{-iH(\lambda_f)t } | \psi(\lambda_i) \rangle,
\end{equation}
where $| \psi(\lambda_i) \rangle$ is the eigenstate of the Hamiltonian $H(\lambda_i)$. The corresponding Loschmidt echo (return probability) is expressed as
\begin{equation}\label{}
  L(t, \lambda_i, \lambda_f) =  | G(t, \lambda_i, \lambda_f) |^2.
\end{equation}

\begin{figure}[!ht]
\centering
\includegraphics[width=3.6in]{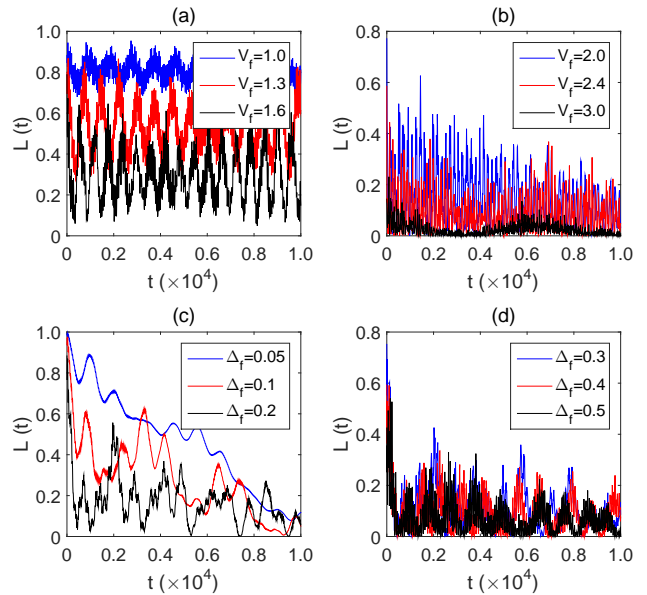}
\caption{(Color online)  The time evolution of Loschmidt echo with different $\lambda_i$s and $\lambda_f$s. (a) and (b) show the variation of $L(t)$ of the AAH model with SC pairing $\Delta=0.2$ after quenches from $V_i=0$ to different $V_f$s. (c) and (d) show the time dependence of $L(t)$ of an AAH model with $\Delta_i=0$ in the initial Hamiltonian and different $\Delta_f$s after a quench. The on-site potential here is $V=2.0$.  The number of the lattice sites is $N=200$.}
\label{fig4}
\end{figure}

With the help of Loschmidt echo, now we can check the phase transition in the AAH model with \textit{p}-wave superconducting pairing from the perspective of dynamical phase transitions. In Figure \ref{fig4}, we present the $L(t)$ of the system after different quenches. In Fig. \ref{fig4}(a) and \ref{fig4}(b), we keep the SC pairing fixed and change the on-site potential. $V_i=0$ in the initial Hamiltonian and the SC pairing is $\Delta=0.2$. So in the beginning, the system is in the extended state. The Loschmidt echo shows very different behaviors when the on-site potential $V_f$ falls into regimes corresponding to the different phases of the system. If the system is in the extended phase after a quench, e.g. $V_f=1.0$ and $1.3$, $L(t)$ oscillates around a relative large value and always keeps finite. It will not touch the zero point. However, when we increase $V_f$ and quench the system into the critical region ($V_f=2.0$, and 2.4), the value of $L(t)$ will decrease and will touch zero at some time intervals. This is the signature of dynamical phase transition. If we increase $V_f$ further (e.g. $V_f=3.0$), the system will go into the localized state, $L(t)$ decreases sharply and will become almost zero in the end. So from the Loschmidt echo, we can also get clear evidences about the three different phases in the AAH model with SC pairing. The critical values for the phase boundaries can be extracted from the results, which are consistent with the ones we get in the phase diagram and the entanglement entropy dynamics in the last section. We also calculate the $L(t)$  after the system is quenched from $\Delta_i=0$ into situations with different SC pairing amplitude $\Delta_f$s, as shown in Fig. \ref{fig4}(c) and \ref{fig4}(d). The on-site potential $V$ here is also set to 2.0, which means that the system is always in the critical phase before and after the quench. If $\Delta_f$ is small (see Fig. \ref{fig4}(c)), it takes a long time before the echo approaches zero. However, if the $\Delta_f$ becomes larger, the Loschmidt echo decreases much faster and will touch zero at some times after the echo oscillates for a shorter time scale. The universal feature that the echo will touch the zero point at certain time intervals always shows up. So from the perspective of Loschmidt echo, the phase transition in the AAH model with SC pairing can also be well characterized.

\section{Summary}\label{sec4}
In this paper, we have investigated the quench dynamics of an Aubry-Andr\'e-Harper model with \textit{p}-wave superconducting (SC) pairing. Due to the presence of the SC pairing, there are three different phases in this system: extended, critical and localized phase. As to the quantum quench, the system is prepared in the extended state with zero on-site potential, and then suddenly quenched into one of the three phases. We check the post-quench dynamical properties of entanglement entropy (EE), mean width of wave packets and Loschmidt echo of this system. These three quantities show distinct but consistent behaviors in different phases, which can be used to characterize the non-equilibrium dynamics of the AAH model. If the system is quenched into the extended states, the EE of the system will increase as a power function of time in the beginning and then saturate. The corresponding Loschmidt echo oscillates around a relatively large value and always keeps finite. However, if the system is in the localized state after a quench, the EE increase quite slowly, and the corresponding Loschmidt echo decays almost to zero in the long-time evolution. Moreover, when the system is quenched into the critical regime, the EE increases in a power-law form and the exponents of the power function are almost the same for different quench values of the on-site potential. The Loschmidt echo in this phase also oscillates at finite value but would reach to zero at some time intervals, which indicates the dynamical phase transitions. The dynamical behaviors of the EE can be partly explained from the perspective of the propagation of wave packets, which exhibits similar dynamical features as the EE in different phases and quench situations. These methods reveal the variations of different physical quantities during the localization-delocalization phase transition process in the AAH model with \textit{p}-wave SC pairing. We also discuss the dynamics of these quantities in the critical phase with various SC pairing amplitudes. Both EE and the mean width of wave packets grow in a power-law form. The exponents of these power functions are almost unchanged for different SC pairings, which indicates the universal property of critical phase. The Loschmidt echo in this case also shows the universal feature that it will reach to zero at certain time intervals.

As for the experimental side, the AAH model can be realized by using cold atom systems. The R\'enyi entanglement entropy, which is similar in characterizing the entanglment as the von Neumann entanglement entropy discussed in this paper, can be measured by using the method proposed in Ref. \cite{Islam}. Moreover, with p-wave SC pairing present in this model, the system is similar to the 1D Kitaev chain, which shows topological superconducting phase with appropriate parameters. The topological phase in the incommensurate 1D AAH model with \textit{p}-wave SC pairing is discussed in Ref. \cite{Cai}, where the topological superconducting phase is found to be destroyed when the system goes into the Anderson localization phase. It would be interesting to further check how the Majorana bound state at the ends of the 1D system evolve with time after quench the system from topological superconducting phase to the localized phase in the future.

\section*{Acknowledgments}
Q.-B. Zeng would like to thank R. Zhang and Y. C. Wang for valuable advices. This work has been supported by the NSFC under Grant No. 11274195 and the National Basic Research Program of China (973 Program) Grant No. 2011CB606405 and No. 2013CB922000. Shu Chen is supported by NSFC under Grants No. 11425419, No. 11374354 and No. 11174360, and the Strategic Priority Research Program (B) of the Chinese Academy of Sciences (No. XDB07020000).

\section*{Appendix}
\setcounter{equation}{0}
\renewcommand{\theequation}{{A}.\arabic{equation}}
\setcounter{figure}{0}
\renewcommand{\thefigure}{{A}\arabic{figure}}

In this Appendix, we present numerical analysis of the time-dependent entanglement entropy (EE) and width of wave packets in the AAH model with \textit{p}-wave superconducting pairing. To better characterize the dynamical behaviors of these quantities, we plot the EE and the mean width of wave packets as a function of time in log-log coordinate systems.

\begin{figure}[!ht]
\centering
\subfigure[]{
\label{figA1a}
\includegraphics[width=3.0in]{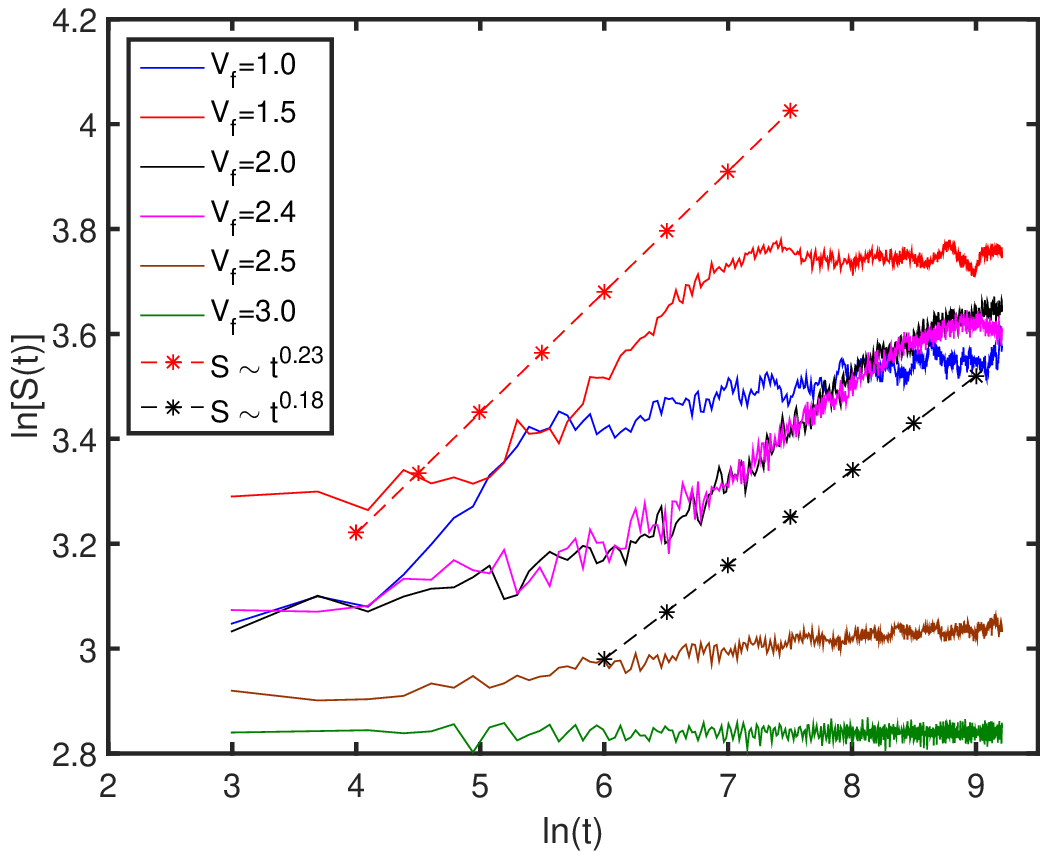}}
\subfigure[]{
\label{figA1b}
\includegraphics[width=3.0in]{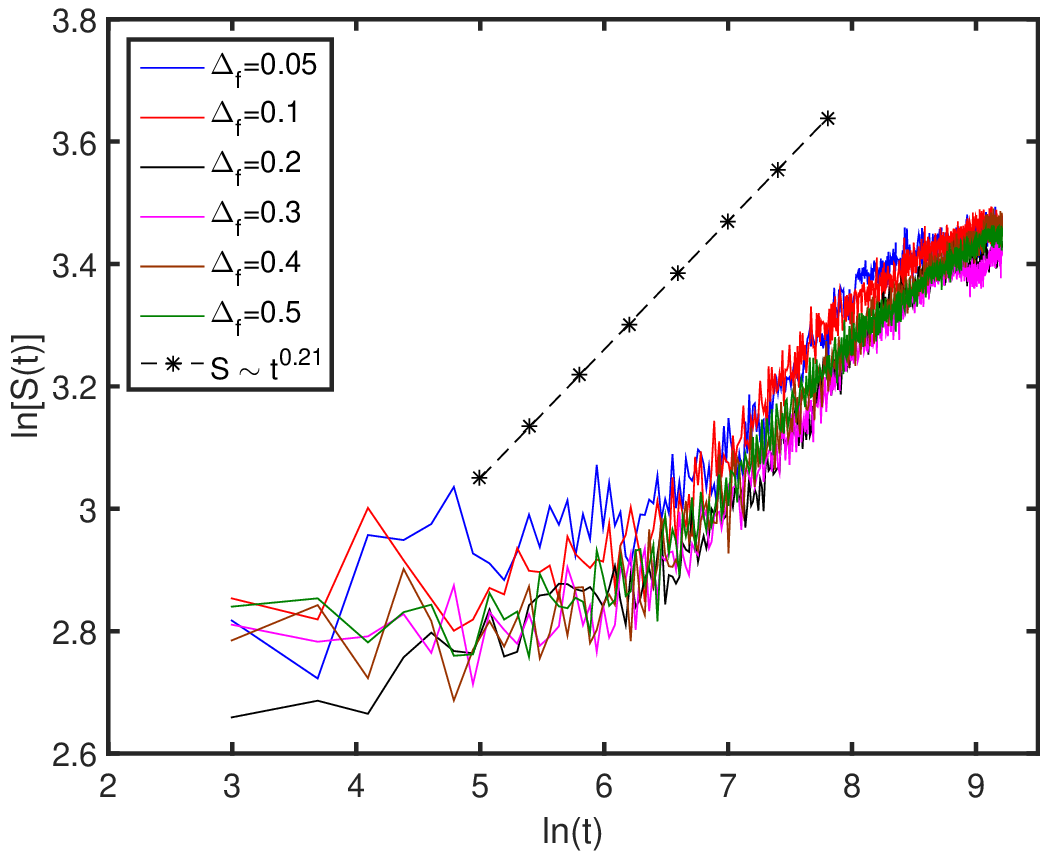}}
\caption{(Color online) The time evolution of the entanglement entropy in the AAH model with \textit{p}-wave superfluidity after a quench. (a) The on-site potential in the model Hamiltonian is set to be 0 before the quench and then it is suddenly changed to different $V_f$s with $1.0 \leq V_f \leq 3.0$ at time $t=0$. Here the SC pairing amplitude is fixed at $\Delta=0.2$. (b) The initial SC pairing amplitude $\Delta_i=0.001$ and then it is suddenly changed to a different value $V_f$ within the range $0.05 \leq \Delta_f \leq 0.5$. The on-site potential is $V=2.0$, so the system is always in the critical regime. The number of the lattice sites is chosen to be $N=233$, which is the 13th Fibonacci number.}
\label{figA1}
\end{figure}

\begin{figure}[!ht]
\centering
\subfigure[]{
\label{figA2a}
\includegraphics[width=3.0in]{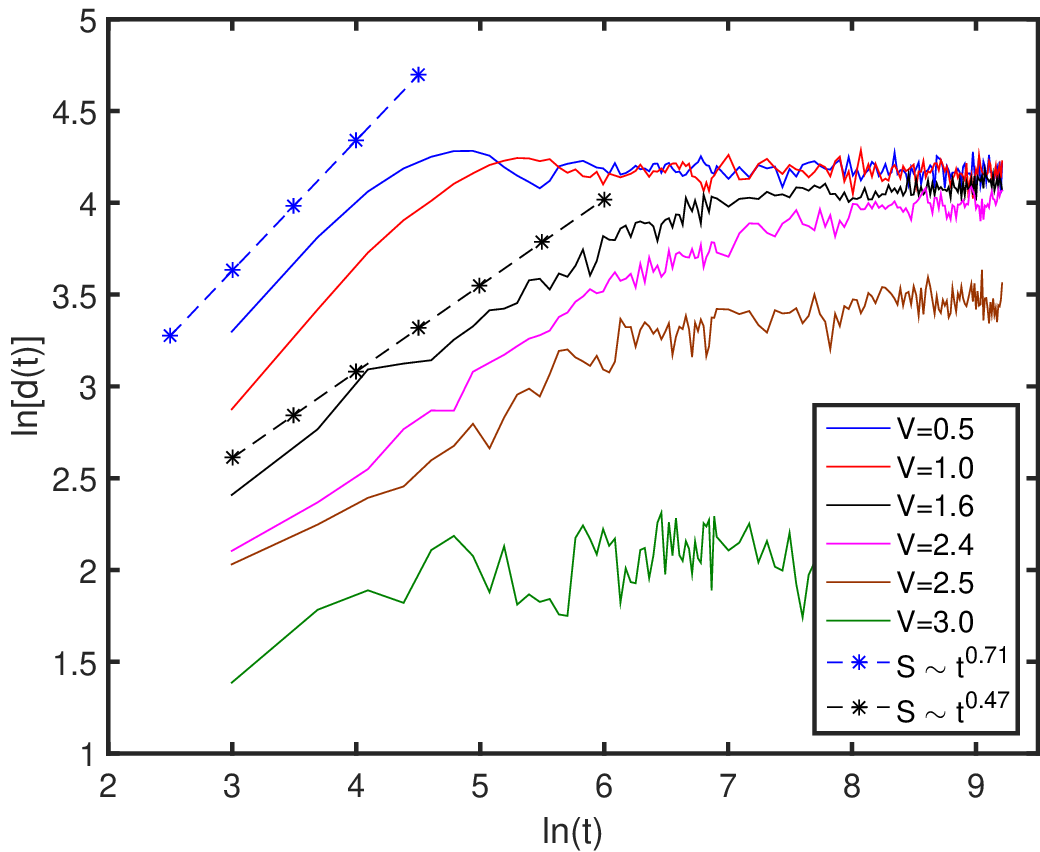}}
\subfigure[]{
\label{figA2b}
\includegraphics[width=3.0in]{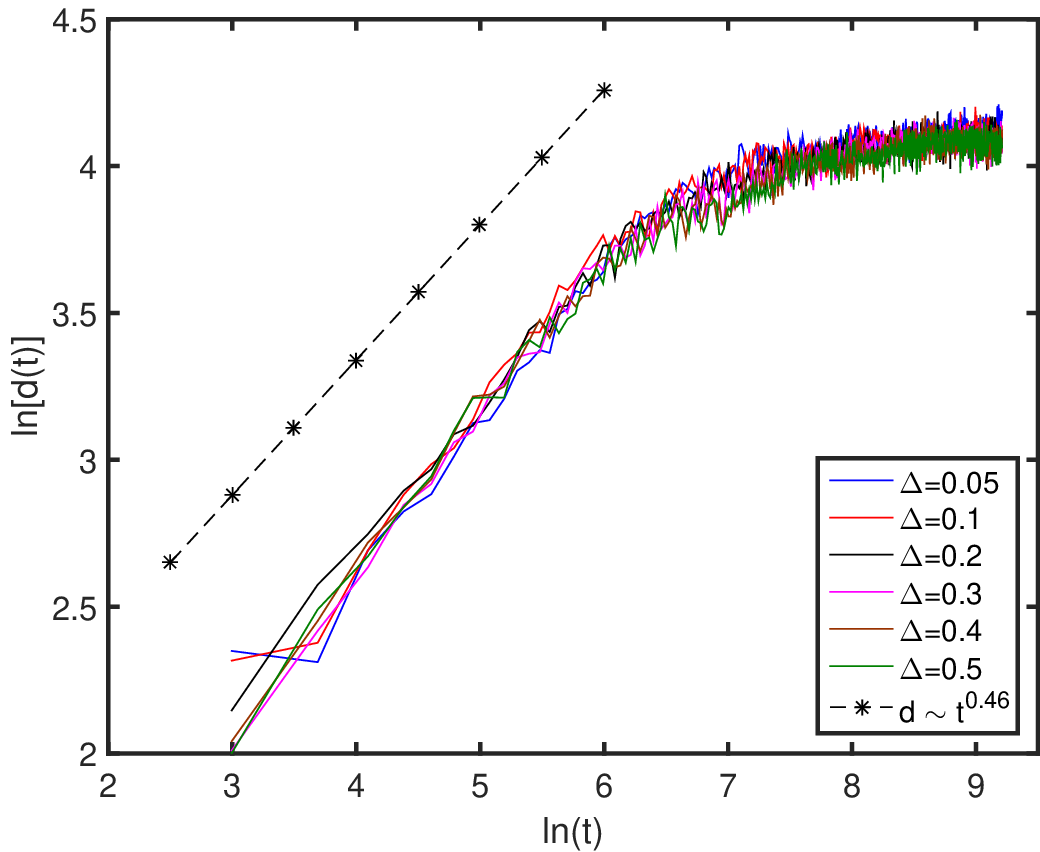}}
\caption{(Color online) Time-dependent width of wave packets in the AAH model with \textit{p}-wave SC pairing. (a) Mean width of wave packets in the systems with different on-site potentials $V=0.5- 3.0$. The SC pairing amplitude is $\Delta=0.2$. (b) Mean width of wave packets in the systems with various SC pairing amplitudes, $\Delta=0.05-0.5$. Here the on-site potential is fixed to $V=2.0$, so the system is always in the critical phase.  The number of the lattice sites is $N=233$.}
\label{figA2}
\end{figure}

Fig. \ref{figA1} shows the time evolution of EE of the system after quench. The growth part of the EE are linear in the log-log frame, which means that the EE grows in a power-law form, $S(t) \sim t^{\sigma}$. This is quite clear for the system after a quench in extended or critical phase. From Fig. \ref{figA1}(a), we can see that $\sigma$ changes as the $V_f$ in the extended phase increases. When $V_f=1.5$, the numerical fitting shows that the exponent $\sigma$ is about $0.23$, see the red dashed line in it. If the system goes into the critical phase, then exponents for various $V_f$ situations are almost the same, $\sigma \approx 0.18$ as indicated by the black dashed line. The features of the critical phase are thus manifested here in the dynamical behavior of the EE by the unchanged exponent. For system quenched in the localized phase, the $\sigma$s are quite small and the EE becomes saturated soon. In Fig. \ref{figA1}(b), we show the time-dependent EE in the system with different SC pairing amplitudes. We have set $V=2.0$ here, so the system is always in the critical phase. It is obvious from this figure that the EE of the critical system with various SC pairings after a quench shows very similar behaviors. The exponent of the power function, $\sigma$, is about $0.21$. This again reveals the signature of the critical phase in the dynamics of the AAH model.

In Fig. \ref{figA2}, we present the time-dependent width of wave packets in the AAH model. The growth of the mean width $d(t)$ also shows a power-law form, $d(t) \sim t^D$. For system in different phases (see Fig. \ref{figA2}(a)), the exponent of the power function will decrease as we increase the value of $V$. When $V=0.5$, $D$ is about 0.71. If the system is in the critical phase, the value of $D$ are almost the same, $D \approx 0.47$, as can be seen from the black and magenta line for $V = 1.6$ and 2.4, respectively. This is consistent with the behavior of EE in the critical phase where similar features also show up. If the system is in the localized phase after a quench, the exponent $D$ would keep decreasing and the mean width of wave packets will become saturated soon. Again we also show the $d(t)$ of system in the critical phase with different SC pairing amplitudes in Fig. \ref{figA2}(b). The power-law growth pattern is obvious and we have $D \approx 0.45$ here. So the dynamical variations of EE and the mean width of wave packets exhibit consistent behaviors in different phases.

{}

\end{document}